# Quantifying Charge Carrier Mobilities and Recombination Rates in Metal Halide Perovskites from Time-Resolved Microwave Photo-conductivity Measurements

*Tom J. Savenije,\* Dengyang Guo, Valentina M. Caselli, and Eline M. Hutter*

Dr. Tom J. Savenije, Dr. Dengyang Guo and Valentina Caselli

Department of Chemical Engineering, Delft University of Technology, 2629 HZ Delft, The Netherlands

Dr. Eline M. Hutter

Center for Nanophotonics, AMOLF, Science Park 104, 1098 XG, Amsterdam, The Netherlands

E-mail: T.J.Savenije@tudelft.nl

**Keywords**: metal halide perovskites, charge carrier dynamics, TRMC, solar cells

**Abstract**

The unprecedented rise in power conversion efficiency of solar cells based on metal halide perovskites (MHPs) has led to enormous research effort to understand their photo-physical properties. In this paper, we review the progress in understanding the mobility and recombination of photo-generated charge carriers from nanosecond to microsecond time scales, monitored using electrodeless transient photoconductivity techniques. In addition, we present a kinetic model to obtain rate constants from transient data recorded using a wide range of laser intensities. For various MHPs the temperature dependence of the mobilities and recombination rates are evaluated. Furthermore, we show how these rate constants can be used to predict the upper limit for the open-circuit voltage $V_{oc}$ of the corresponding device. Finally, we discuss photo-physical properties of MHPs that are not yet fully understood, and make recommendations for future research directions.

**Introduction**

The power conversion efficiency of metal halide perovskite (MHP) solar cells has increased uninterruptedly since 2009, reaching over 25% in 2019.[1–4] This extraordinary development of MHP semiconductors has changed the landscape of solar cell research prompting many groups to refocus their efforts to this emerging field. Recently, it has been shown that MHPs can also be used in light emitting diodes, lasers, and even in radiation detectors.[5–10] In this paper, we present an extensive overview of the dynamics of light-induced charge carriers in MHP semiconductors. We show how directly monitoring the changes in the conductivity of the material on optical excitation using electrodeless microwave techniques enables to determine both the mobility and recombination lifetime of free charges.

In the first section of this paper, we present the technical details of the electrodeless time-resolved microwave conductivity (TRMC) technique. Section two introduces a kinetic model to obtain from the TRMC results quantitative data, including charge carrier mobilities and decay constants. These mobilities will be compared to other reported values determined by *e.g.* terahertz spectroscopy. The third section discusses the temperature-dependent mobility and recombination rates for a number of different MHPs. The fourth section shows that if the rate constants describing the dynamics are accurately known, it is possible to predict the open circuit voltage of a corresponding solar cell. In the final part we show that for a number of MHPs, including mixed cation perovskites, the model presented in Section 2 fails to describe the charge carrier dynamics. More elaborate models have to be developed for these systems which will teach us more about the band structure and intra-band gap levels of these more complex perovskite semiconductors.

**Section 1: Time-Resolved Microwave Conductivity**

The TRMC technique can be used to study the dynamics of photo-induced charge carriers in semiconductor materials with low background conductivities.[11–15] This technique is based on the interaction between the electric field component of microwaves (GHz regime) and mobile carriers. Hence, the photoconductivity can be determined without contacting the semiconductor to electrodes, thereby avoiding interfacial effects or undesired chemical reactions between the MHP and the metallic electrodes.[16,17]

In general, if photo-excitation of a material results in the generation of free, mobile charge carriers, the conductivity of this material is enhanced. By definition, the electrical conductivity, $\sigma$ scales with the concentration of free charge carriers, $n$ and their mobility $\mu$ according to:

$$\sigma = e \sum_i n_i \mu_i \qquad (1)$$

where $e$ is the elementary charge. With the TRMC technique, the change in conductivity between dark and after illumination, *i.e.* $\Delta\sigma$, is probed. Although optical excitation can result in a gradient in the conductivity throughout the sample, TRMC records the integrated change in $\Delta\sigma$ over the complete film thickness, which yields the change in conductance, $\Delta G$:

$$\Delta G = \beta \int_0^L \Delta\sigma(z)\, dz \qquad (2)$$

Here $\beta$ is the ratio between the wide and small inner walls of the microwave guide. Hence, $\Delta G$ is proportional to the product of the total number of photo-induced free charges and their mobility. Absorption of microwaves by photo-induced charges reduces the microwave power, $P$ on the detector. $\Delta P$ is recorded as function of time after the laser pulse and thus, the TRMC technique can be used to determine both the mobility and time-dependent concentration (*i.e.*

the lifetime) of photo-induced free charges. This is schematically depicted in Figure 1. The normalized reduction in microwave power $\frac{\Delta P(t)}{P}$ is related to $\Delta G(t)$ by:

$$\frac{\Delta P(t)}{P} = -K\Delta G(t) \qquad (3)$$

where $K$ is the sensitivity factor, $\Delta P(t)$ is negative while $\Delta G(t)$ is positive. Quantification of the $K$-factor was previously described.[15,18]

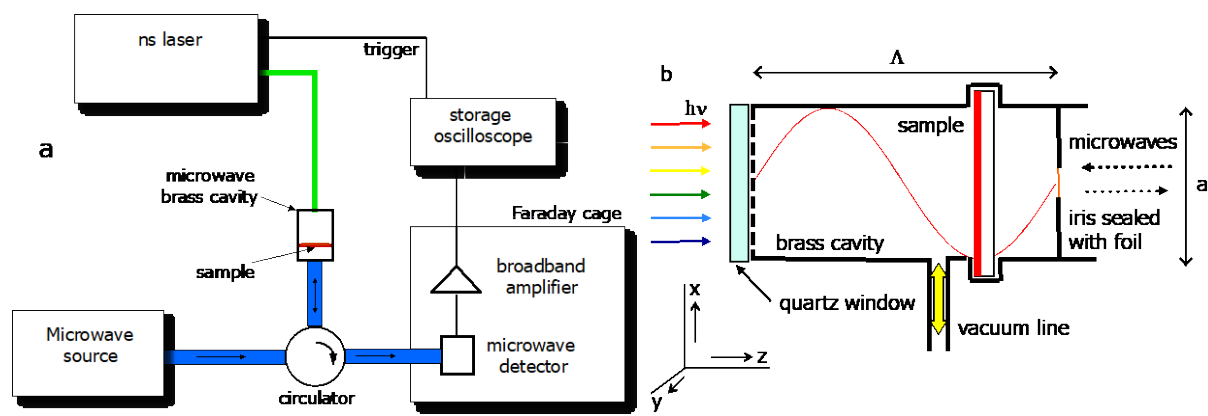

**Figure 1.** (a) Schematic representation of the TRMC set-up. Monochromatic microwaves are generated using a voltage-controlled oscillator (microwave source). The sample of interest is placed in a microwave cell (b), at approximately ¾ of cell length $\Lambda$ to maximize overlap with the electric field of the microwaves with wavelength $\Lambda$. A circulator separates incident from the reflected microwaves. Reprinted with permission from Savenije *et al.*[13] Copyright 2013 American Chemical Society.

Figure 1 shows a representation of the TRMC set up. Photo-excitation is realized by a laser pulses of 3-5 ns FWHM with a tunable wavelength at a repetition rate of 10 Hz. Metallic, neutral density filters with different optical densities are placed in between the laser and the sample to vary the photon fluence between $10^9$ and $10^{15}$ photons/cm$^2$ per pulse. Monochromatic microwaves with a frequency in the range from 8.2 to 12.4 GHz are generated using a voltage-controlled oscillator. The sample of interest is placed in a microwave cell that ends with a metal grating (see Figure 1b), which largely transmits the laser excitation source while fully reflecting the microwaves. This cell is made from a gold-plated X-band waveguide. The sample is placed at ¾ of the cell length, $\Lambda$ so that its position corresponds to the maximum electric field strength for microwaves with wavelength λ. A quartz window is glued on top of the grating to seal the cell and avoid air exposure of the sample. The cell is connected to a continuous 100 mWatt microwave source and detector via X-band waveguides, see Figure 1a. A microwave circulator is incorporated to separate the incident from the reflected microwaves. The diode detector converts the microwave power into a direct current (DC), generating a voltage of 0.5-1.0 V when dropped across a 50 Ohm resistor. Since, the laser induced change in microwave power, $\Delta P(t)$ is several orders of magnitude smaller than $P$, an offset regulator is used to subtract the DC part, which leaves the laser induced AC signal undisturbed. Then the AC signal is amplified by a broadband amplifier and stored as function of time using a digital oscilloscope with a gigahertz sampling rate. The trigger input of the oscilloscope is connected to a fast optical sensor that is illuminated by each laser pulse to start the acquisition of a microwave trace. Typically, the TRMC traces are averaged over 10 - 1000 pulses. If the sensitivity $K$ is known, $\Delta G$ can be quantitatively obtained from the measured $\Delta P(t)$ /P using Eq. 3. Then, the TRMC signal can be expressed in the remaining two unknown parameters: *i.e.* the mobility and the charge carrier yield. If every absorbed photon creates a single positive and a negative charge carrier, Eq. 1 simplifies to:

$$\sigma = en(\mu_e + \mu_h) \qquad (4)$$

in which $n$ is the concentration, $\mu_e$ is the electron mobility and $\mu_h$ is the hole mobility. The yield, $\varphi$, can be defined as

$$\varphi = \frac{Ln}{F_A I_0} \qquad (5)$$

in which $I_0$ is the intensity of the laser in photons/pulse/unit area and $F_A$ the fraction of light absorbed at the excitation wavelength. By combining Eqs. 2, 4 and 5, the product of yield and mobility can be obtained from $\Delta G_{max}$

$$\varphi(\mu_e + \mu_h) = \frac{Ln}{F_A I_0}\frac{\Delta\sigma}{en} = \frac{L}{F_A I_0}\frac{\Delta G_{max}}{e\beta L} = \frac{\Delta G_{max}}{F_A I_0 e\beta} \qquad (6)$$

Expressing the TRMC signal in $\varphi(\mu_e + \mu_h)$ product enables us to directly compare TRMC measurements of different samples.

The sensitivity of the TRMC set up can be further increased by partially closing the cell using an iris (see Figure 1b). In this case, the cell acts as a resonant cavity for microwaves with wavelength $\lambda$. The standing wave pattern in the cavity leads to more interaction with the sample, thereby enhancing $K$ and enabling the use of lower photon fluences, however at the expense of a decreased time resolution. For a cavity with an approximately 40 times increased $K$ value, the instrumental response time is close 18 ns, while this is only ca 1 ns when a measurement is performed with the open cell. In contrast to DC techniques, such as time-of-flight or field effect transistor measurements, charges do not undergo a net displacement in the photo-active layer during a TRMC measurements. In fact, due to the low electric field strength of the microwaves in the cell and the rapid oscillating direction of the electric field, the drift distances are relatively small, *i.e.* in the nanometer regime, see also Figure 2.

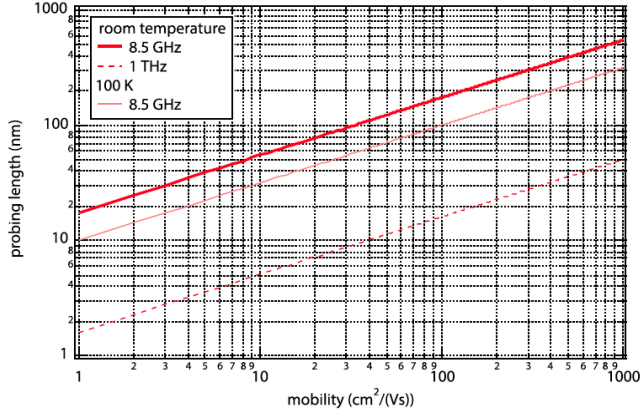

**Figure 2**. Relationship between the measured mobility and probing length (nm) in time-resolved photoconductance measurements. Here, the probing length represents the distance over which charges diffuse at probe frequencies of 8.5 GHz (microwave conductivity, solid lines) or 1 THz (terahertz spectroscopy, dotted line). Adapted from Muscarella *et al*.[19]

Therefore, one can envision the movement of the light-induced charge carriers in the sample by their diffusional motion, slightly perturbed by the microwave electric field. It is important to note here that if the diffusional motion is limited by the sample boundaries the resulting TRMC mobility is smaller than the ultimate DC mobility observed in *e.g.* a single crystal.[20] Figure 2 illustrates the probing length as a function of the DC mobility and the used frequency.[19] For MHPs with a typical measured mobility of 20 cm$^2$/(Vs) the probing length is of the order of 50 nm, while for terahertz spectroscopy measurements the probing length is one order of magnitude lower. In practice, this means that the TRMC signal measured in crystallites substantially smaller than 50 nm, like MHPs grown within a mesoporous scaffold,[21] the effective observed mobility values are lower than that in a DC experiment. However, the mobilities observed with TRMC are often substantially larger than those determined by a time-of-flight or field effect transistor measurement. This is due to the fact that in a DC experiment charges have to cross many grain boundaries in a polycrystalline layer, which will lower the measured mobility substantially.[22]

Considering that both electrons and holes contribute to the photo-conductance, $\Delta G(t)$ is proportional to the sum of both concentrations and their mobilities. This is in contrast with more frequently used time-resolved photoluminescence (PL) measurements, which specifically detects radiative recombination events. Therefore, the decays obtained with TRMC cannot a priori directly be compared to PL transients. Instead, TRMC gives complementary insights to time-resolved PL, since it probes the recombination of all free mobile charge carriers: both radiative and non-radiative, see Figure 3. Therefore, TRMC is especially useful for studying the recombination kinetics in MHPs showing poor light-outcoupling and substantial reabsorption of emitted photons, such as macroscopic MHP crystals, or MHP-transport material bilayers, where most recombination is non-radiative.

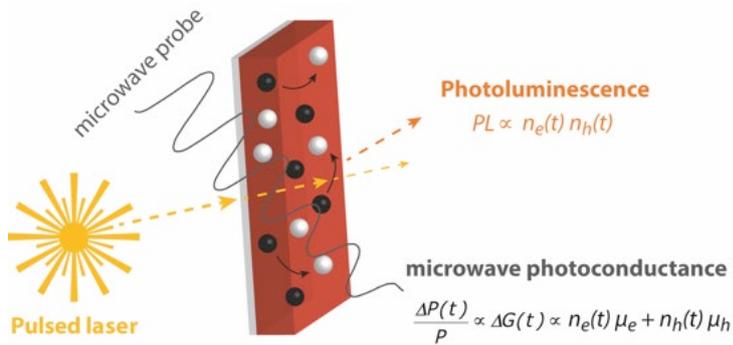

**Figure 3.** Representation of time-resolved photoluminescence (TRPL) and TRMC measurements. In both techniques, electrons (black spheres) are excited to the conduction band by a short laser pulse, yielding mobile holes (white spheres) in the valence band. TRMC is used to measure the photo-conductance ($\Delta G(t)$), which scales with the time-dependent concentration $n$ and mobility $\mu$ of free electrons $e$ and holes $h$. The sinusoidal line represents the magnitude of the microwave electric field as it passes through the sample. Radiative recombination

of these mobile electrons and holes can be probed by TRPL, which is a function of the concentrations of electrons and holes.

**Section 2: Modelling of Kinetic Data**

In order to extract quantitative data out of TRMC measurements, the following generic kinetic model can be used, as detailed in Figure 4.[21,23] This model accounts for different recombination pathways of photo-excited electrons and holes and assumes homogeneous generation and decay of charges, which can be experimentally realized by using an excitation wavelength close to the absorption onset. In view of the relatively small exciton binding energy compared to thermal energy in most bulk MHPs,[24] electron-hole pairs dissociate into free CB electrons and VB holes. Note that only free, mobile charges contribute to the real photo-conductance.

Formation of charge carriers is represented by the generation term, $G_C$ and takes into account the temporal profile and total light intensity of the laser pulse. In general, semiconductors are often unintentionally doped due to impurities in the crystal lattice. This doping leads to additional CB electrons (n-type) or VB holes (p-type) already present before photo-excitation. In Figure 4, the concentration of thermal equilibrium charges (dark carriers) is represented by $p_0$. Note that $p_0$ does not contribute to the photo-conductance. However, the recombination of photo-generated charges is affected by $p_0$, because the total concentration of VB holes ($n_h + p_0$) is larger than the concentration of CB electrons ($n_e$).

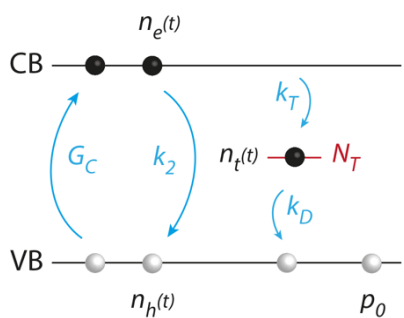

**Figure 4.** Kinetic model of processes occurring upon photo-excitation of a MHP. Here, $G_C$ represents the photo-excitation of electrons (closed circles) from the valence band (VB) to the conduction band (CB). The CB electrons can recombine with holes (open circles) in the VB via $k_2$. In competition with $k_2$, electrons can be immobilized in intra-band gap trap states (with total density $N_T$) via $k_T$. Finally, the trapped electrons can recombine with holes from the VB via $k_D$. In the case of an (unintentionally) doped MHP, there will be additional holes ($p_0$, p-type) on top of the photo-generated holes. Note that this fully mathematical model also holds for the opposite situation, *i.e.* an MHP with trap states for holes and additional dark CB electrons ($n_0$, n-type).

The following set of differential equations implements Figure 4 and describes $n_e$ (Eq. 8) $n_h$ (Eq. 9) and $n_t$ (Eq. 10) as function of time. The rate constants for band-to-band electron-hole recombination, trap filling and trap emptying are represented by $k_2$, $k_T$ and $k_D$ respectively.

$$\frac{dn_e}{dt} = G_c - k_2 n_e(n_h + p_0) - k_T n(N_T - n_t) \qquad (8)$$

$$\frac{dn_h}{dt} = G_c - k_2 n_e(n_h + p_0) - k_D n_t(n_h + p_0) \qquad (9)$$

$$\frac{dn_t}{dt} = k_T n_e(N_T - n_t) - k_D n_t(n_h + p_0) \qquad (10)$$

Solving the equations using numerical methods yields the time-dependent concentrations $n_e$, $n_h$ and $n_t$. The change in photo-conductance as function of time is calculated from the product of charge carrier concentrations and mobilities according to:

$$\Delta G(t) = e(n_e(t)\mu_e + n_h(t)\mu_h)\beta L \qquad (11)$$

Here, the trapped charge carriers, $n_t$ are assumed to be immobile and do not contribute to $\Delta G$. The mobilities are assumed to be constant within the time window of the measurement and independent of the charge density. Finally, a convolution is applied to take into account the instrumental response function of the set-up.[15]

## Section 3: TRMC measurements on MAPbI$_3$

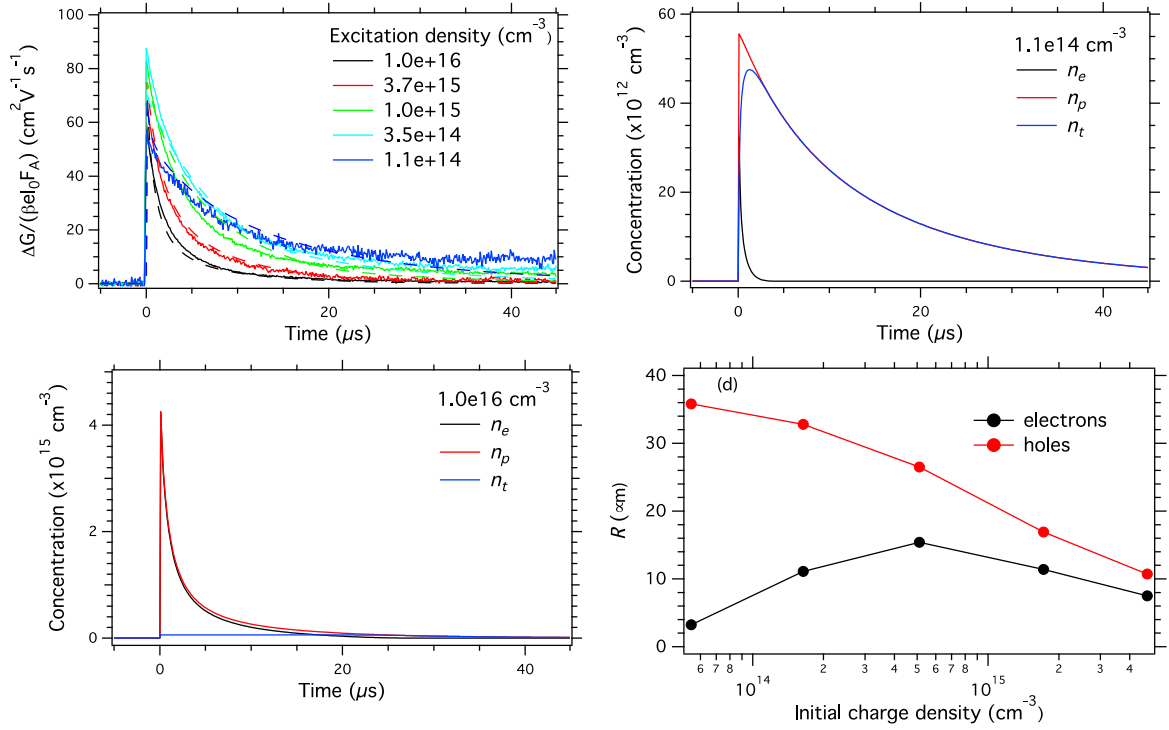

**Figure 5.** Experimental TRMC traces for (a) MAPbI$_3$ recorded at different incident light intensities ($\lambda$ = 600 nm, 10 Hz), corresponding to initial charge carrier densities of $10^{14}$ to $10^{16}$ cm$^{-3}$, adapted from Ref.[25] The dashed lines show the traces calculated by solving the differential equations (8 - 10) and converting the time-dependent concentration curves to normalized $\Delta G$. Panels (b) and (c) show the calculated concentrations for an initial excitation density of $10^{14}$ and $10^{16}$ cm$^{-3}$, respectively. Panel (d) shows the calculated charge carrier diffusion lengths on basis of charge carrier lifetime and mobility.

A typical example of TRMC measurements on MHPs is provided in Figure 5. In panel (a), the photo conductance, normalized for the number of incident photons is shown as function of time on photo-excitation at $\lambda$ = 600 nm using incident light intensities ranging over two orders of magnitude. The rise of the signal at $t$ = 0 originates from the formation of mobile excess charges; the maximum signal height represents the product of charge carrier generation yield and the mobility, see also Eq. 6. At an excitation density of $10^{14}$ cm$^{-3}$, the yield-mobility product $\varphi(\mu_e + \mu_h)$ amounts to 58 cm$^2$V$^{-1}$s$^{-1}$, whereas $\varphi(\mu_e + \mu_h)$ = 87 cm$^2$V$^{-1}$s$^{-1}$ at higher excitation densities.[25] We interpret this as a reduction in the yield of free charges at low excitation densities, due to rapid immobilization in defect states. Slow recombination of these immobilized charge with free charges could then lead to relatively long lifetimes. At higher intensities, the decay kinetics become increasingly faster, which is typically observed in a regime where higher order recombination dominates. Using the kinetic model as described in Section 2, we can fit the TRMC traces using a single set of kinetic parameters (see Table 1). For obtaining the different TRMC traces, only the generation term $G_c$ is altered according to the incident intensity. We see that the model describes the kinetic traces very well, in particular the reduction in signal size at low excitation densities *i.e.* a photo-excited charge carrier density lower than $N_T$. Since at those intensities all minority carriers get immediately trapped in deep trap states, the electrons have a negligible contribution to $\varphi(\mu_e + \mu_h)$, which enables us to selectively derive $\mu_h$. This fast trapping process occurs until at higher excitation densities all available trap states are occupied. At even higher excitation intensities the effect of trapping states can be neglected and signal is dominated by $k_2$.

The above discussion can be visualized by plotting the concentrations of $n_e$, $n_h$ and $n_t$ separately as shown in Figures 5b and 5c for excitation densities of 10$^{14}$ cm$^{-3}$ and 10$^{16}$ cm$^{-3}$, respectively. For low excitation densities (Figure 5b) the electrons, which we assume to be the minority carriers in this sample are rapidly trapped in deep trap states. Therefore, at low excitation densities, the TRMC mainly probes the decay of mobile holes. As shown in Figure 5b, this is dominated by the emptying of trapped electrons back to the VB, as clearly shown by the evolution of $n_t$. At higher intensities, see Figure 5c, the excess carrier concentrations exceeds $N_T$ and $p_0$, yielding approximately identical decay curves for $n_e$ and $n_h$. From the concentration curves it is feasible to extract the half-life times, $\tau_{1/2}$ to deduce the charge carrier diffusion length, $R$ for the minority and majority carriers using

$$R = \sqrt{D\tau_{1/2}} \qquad (12)$$

The electron and hole diffusion coefficients, $D_e$ and $D_p$ can be derived from the electron and hole mobilities obtained by fitting of the TRMC traces and using the Einstein-Smoluchowski equation $D = kT\mu/e$. As shown in panel (d) at low intensities $R$ for the minority carriers is small in contrast to that of the majority carriers. At excitation intensities yielding charge carrier concentration close to $N_T$ we observe the maximum values of $R$ for the electrons. At higher densities $R$ becomes smaller for both types of carriers due to enhanced band-to-band recombination ($k_2$). Thus, the electron and hole diffusion lengths are most balanced at charge densities just above the trap density, which is desirable for optimum collection of both electrons and holes in an operating solar cell.

As mentioned above, different experimental techniques can be applied to determine the mobilities in perovskite materials, including AC techniques like time-resolved microwave conductivity and THz spectroscopy,[14,21,26–29] and DC techniques such as Hall effect,[30,31] and space charge limited current (SCLC) measurements.[32] A direct comparison of the experimentally determined values is not always possible, due to differences in the sample quality and preparation, but also due to excitation conditions. An extensive discussion over the latter is reported for thin films of MAPbI3 by Levine et al.[33] As for single crystals, a proper determination of the carrier mobilities can be even more challenging due to the size of the crystals themselves, as well as intrinsic limiting factors of the experimental techniques, as pointed out by Herz.[34] Among others, the microwave conductivity technique offers advantages both in terms of reliability, sensitivity and reproducibility. TRMC is a contactless technique and thus, it does not require prolonged exposure to electric fields that can potentially lead to ion migration.[35] Additionally, photogeneration can occur over a broad range of intensities in the 1-sun equivalent regime.[33] In Table 1 we report an overview of some experimentally determined mobility values in MAPbI3 thin films and single crystals. Basically, mobility values found by THz and TRMC are within experimental errors. For MAPbI3 films analysis of mobilities yields substantially lower values than that for single crystals, probably due to the polycrystalline nature of the films. For MAPbI3 single crystals on the other hand, mobility values measured by AC or DC techniques are similar. Hence, reducing the unfavorable effect of grain boundaries in MAPbI3 films should lead to long-range mobilities comparable to the (short-range) TRMC mobility.

Table 2: Experimental mobility values for different MAPbI$_3$ thin films and single crystals.

|  | Technique | Mobility (cm$^2$V$^{-1}$s$^{-1}$) | Reference |
|---|---|---|---|
| Thin Film | THz | ~20 (Σ), ~12.5 (e), ~7.5 (h) | [27] |
|  | THz | 35 (Σ) | [36] |
|  | TRMC | 71 (Σ) | [37] |
|  | TRMC | 29 (Σ) | [26] |
|  | TRMC | 87 (Σ) | [25] |
|  | TRMC | ~80 (Σ) | [38] |
|  | TRMC | 30 (Σ) | [21] |
|  | TRMC | 30.5 (e), 6.5 (h) | [39] |
|  | TRMC | 18 (e), 2.5 (h) | [33] |
|  | Hall | 8 | [40] |
| Single Crystal | SCLC | 164 ± 25 (h) | [20] |
|  | Hall | 105 ± 35 (h) |  |
|  | ToF | 24 ± 6.8 (e) |  |
|  | TRMC | 115±15 (Σ) | [41] |
|  | Hall | 11 | [40] |
|  | SCLC | 67.2 | [42] |

(Σ) denotes sum of electron and hole mobility

**Section 4: Temperature dependence of charge carrier dynamics in MHPs**

In this Section we will compare how temperature affects the dynamic processes in MAPbI$_3$ and CsPbI$_3$. The CsPbI$_3$ films are prepared using physical vapor deposition, yielding the metastable black-phase CsPbI$_3$.[43,44] Figure 6 shows intensity-normalized TRMC traces for excess charge carrier densities in the regime between $10^{15}$ and $10^{17}$ cm$^{-3}$. For both samples two observations are evident: first, with lowering the temperature the maximum TRMC signal heights become increasingly higher and secondly, the decay kinetics become slower. In all cases we avoided phase transitions, as these lead to dramatic changes in the crystal lattice.

We used our kinetic model described in Section 3 to analyze the data and the resulting fits are added in Figure 6. It is assumed that for the used temperature range all absorbed photons lead to the formation of free charge carriers, due to their relatively low binding energies,[45] which was further confirmed by electron accelerator measurements.[46] In Figure 7a we plot the mobilities used for the fits as function of temperature. For both samples we see mobilities in the order of several tens (cm$^2$/(Vs)) comparable with values in the literature for these materials.[34] On decreasing the temperature, the mobilities in both MAPbI$_3$ and CsPbI$_3$ gradually increases, following a temperature-dependent trend given by $\mu \propto T^{-\alpha}$, with $1.3 < \alpha < 1.7$.[38] This shows that the charge carrier mobility in lead iodide perovskites is mainly limited by phonon scattering, and is basically independent of the monovalent cation.[44,47]

At present, a debate exists which type of phonons (optical or acoustic) limits the charge mobility in MHPs.[48,49] While optical phonons, leading to the formation of large polarons would predict the experimentally found mobilities much better, the temperature dependence of the mobility is described more accurately by acoustic phonons.[50] That is, acoustic phonons predict that $\mu \propto T^{-1.5}$, whereas for polarons this is $T^{-0.5}$.[48,51,52] Since we and others have found that for many

MHPs $1.3 < \alpha < 1.7$,[28,29,36,53–55] scattering with acoustic phonons seems to be the dominant factor determining mobility instead of scattering with optical phonons.

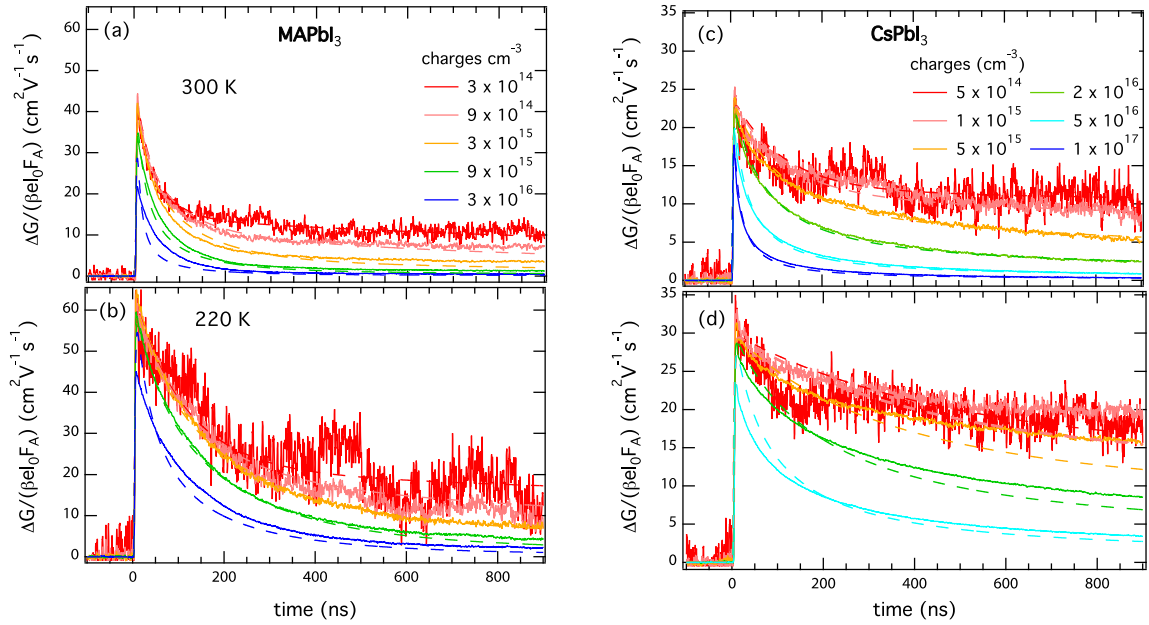

**Figure 6.** Time-resolved photoconductance traces normalized to the absorbed number of photons for spin-coated MAPbI$_3$ (a,b: left panels) and vapour-deposited CsPbI$_3$ (c,d: right panels), recorded at 300 K (a,c: top panels) and 220 K (b,d: bottom panels). The modelled traces are added as dashed lines. Adapted from Hutter *et al.*[47]

Interestingly, not only the mobility but also the lifetime of mobile charges follow the same temperature dependence in MAPbI$_3$ and CsPbI$_3$, as shown in Figure 7b. That is, for a range of excitation densities between $10^{15}$ and $10^{17}$ cm$^{-3}$, the lifetime is enhanced on lowering the temperature. More specifically, although the absolute values are different, in both cases the $k_2$ at 220 K is less than half its room-temperature value. This is in contrast with the temperature-dependent trend in recombination rate typically found in direct bandgap semiconductors and previously reported for MAPbI$_3$ at high fluence (>$10^{17}$ cm$^{-3}$),[36] which is several orders of magnitude higher than the charge densities presented here. Although these previous reports suggest that a direct recombination pathway dominates at higher charge densities, our results show that in both CsPbI$_3$ and MAPbI$_3$, second-order recombination is actually a thermally activated process for charge densities ranging from $10^{15}$ to $10^{17}$ cm$^{-3}$.

Recently, Richter *et al.* suggested that the bimolecular recombination constant is comprised of a radiative ($k_R$) and non-radiative ($k_n$) component.[56]

$$k_2 = k_R + k_n \qquad (13)$$

Here it is important to note that the effective bimolecular recombination rate constant, $k_2$ that can be extracted from our fits to the TRMC data might differ from the values found by *e.g.* time-resolved PL. In the latter technique the radiative process is assessed, while with TRMC we probe by nature the presence of all excess, mobile carriers. Since typically the PL intensity increases with lowering the temperature,[57,58] we speculate that apart from an increasement of $k_R$, $k_n$ decreases dramatically with lower temperatures.[25] Previously, we hypothesized that the non-radiative decay is due to the conduction band minimum (CBM) being slightly shifted in *k*-space with respect to the valence band maximum (VBM) due to Rashba splitting,[54] resulting in an indirect bandgap from which recombination is momentum-forbidden.[59–61] The high absorption coefficients and the temperature-dependent recombination at high charge densities on the other hand indicate the presence of a direct transition.[62,63] The origin of the thermally enhanced non-radiative recombination rates may therefore be twofold: (1) thermal energy releases electrons from the CBM to a state from which direct recombination is possible and (2) the electrons decay from the CBM to the VBM via indirect recombination on interacting with phonons. Other scenarios to explain the thermally enhanced recombination include charge immobilization into shallow traps or the formation of large polarons. However, because both of these do not satisfactorily explain the experimental temperature-dependent mobility following T$^{-1.5}$, these are in our opinion unlikely to dominate the charge carrier recombination properties.

We realize that other explanations for this thermally activated recombination of free charges are still possible, such as an inhomogeneous energy landscape due to local variations in composition or band bending near the surface.[64,65] In addition, the increased lifetime observed upon lowering the temperature could partially be related to an increase in $k_R$ and consequently, increased reabsorption of emitted photons.[56,66,67] However, we note that the external photoluminescence quantum efficiency (PLQE) in $CsPbI_3$ is less than 0.05%,[44,68] so that even in the case of very poor output coupling (*e.g.* 10%) and a 10-fold increase in PLQE on lowering the temperature, still more than 95% of the second-order recombination is non-radiative and therefore cannot be reabsorbed. Hence, photon reabsorption alone cannot explain the thermally activated non-radiative second order recombination that we have observed in $CsPbI_3$ and $MAPbI_3$. Therefore, this intriguing behavior requires more research to fully understand, aiming to design MHPs with long-lived mobile charges available for collection.

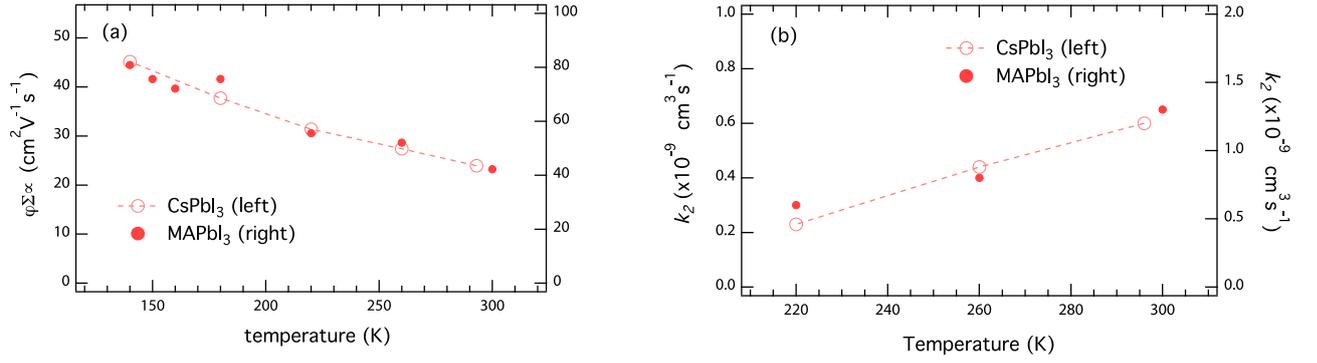

**Figure 7.** Temperature-dependence of (a) the $\varphi(\mu_e + \mu_h)$ product and (b) the effective bimolecular recombination $k_2$. The dashed lines are added to guide the eye. Adapted from Hutter et al.[47]

**Section 5: Influence of charge carrier dynamics on the device performance of MHP solar cells**

In this section we show how we can predict device parameters *i.e.* the open circuit voltage, $V_{oc}$ of an MHP cell using the kinetic data extracted from the time dependent photoconductance measurements on a bare MHP layer. Although the $V_{oc}$ of MHP-based solar cells has increased from 0.61 V[1] to over 1.2 V,[69] this value is still below the $V_{oc}$ that should be feasible based on the bandgap and thermal radiation or so-called entropy losses (1.33 V for MAPbI3).[70–73] Hence, understanding the factors governing the $V_{oc}$ and its improvement is essential to exploit the full potential of MHPs. Since optimizing solar cells is labour-intensive, prediction of the upper limit of the $V_{oc}$ on basis of the characteristics of a bare perovskite semiconductor layer is extremely useful. The product $qVoc$ equals the Quasi-Fermi level splitting, $\mu_F$ under illumination at open circuit and is defined by:[74,75]

$$\mu_F = \frac{kT}{e} ln \frac{(n_0 + n_e)(p_0 + n_p)}{n_i^2} \qquad (14)$$

where the kT/q is the thermal energy, $n_i$ is the intrinsic carrier concentration, $n_0$ and $p_0$ are thermal-equilibrium concentrations of electrons and holes respectively, and $n_e$ and $n_p$ are the concentrations of photo-excited excess electrons and holes, respectively. The latter values can be derived from *e.g.* photoluminescence measurements but also from the kinetic data found by globally fitting the TRMC data, as presented in Section 2. This set of parameters enables us using the differential equations given by Eqs. 6 to 8 to determine the excess electrons and holes as function of time for different generations terms $G_C$. While for fitting the TRMC data we use a generation term representing the intensity and laser profile of the laser pulse, for determining the excess carrier concentrations under continuous solar irradiation, we use a generation term representing AM1.5. On numerical solving the set of differential equations the excess densities can be extracted as shown in Figure 8 illustrating the effect of the incident intensity.

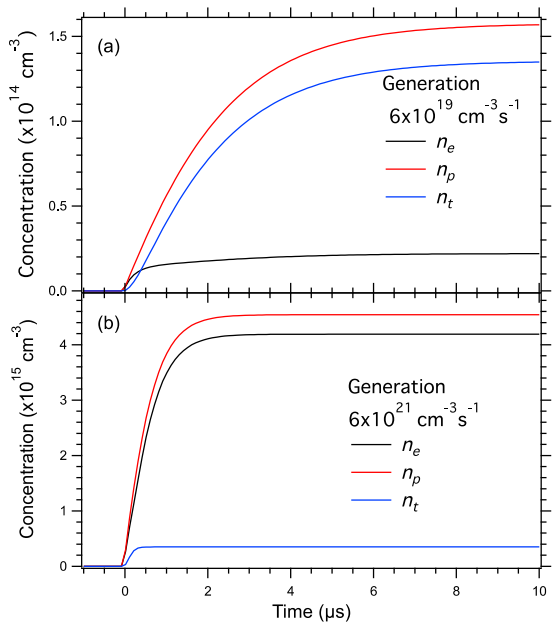

**Figure 8.** Calculated $n_e$ (black), $n_p$ (red), and $n_t$ (blue) under continuous excitation using rate constants determined from the pulsed photoconductance measurements. The steady state excitation densities correspond to (a) 1% and (b) 100% of the number of photons that the MHP layer would absorb under AM1.5.[74]

Knowing $n_e$ and $n_p$ enables us to calculate $\mu_F/e$ using Eq. 14, which is shown for a number of samples given in Figure 9.

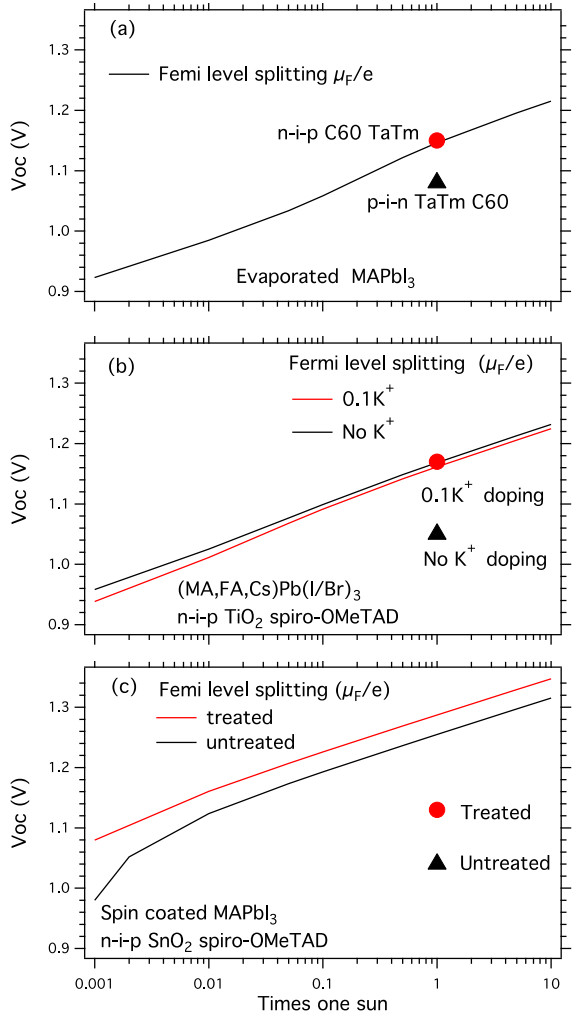

**Figure 9**. Comparison between the calculated Quasi-Femi level splitting (full lines), $\mu_F/e$, and the corresponding experimental $V_{oc}$ (markers). The data shown in (a), (b) and (c) are taken from references [25,76,77], respectively. The deposition method of the MHP layers, device structures and transport materials are given in the annotations. Adapted from Guo *et al*.[74]

Figure 9 shows that although all solar cells are fabricated using the same evaporated MAPbI$_3$, the $V_{oc}$ for the n(C60)-i-p(TaTm) structure is 1.15 V, while for the reverse stack the $V_{oc}$ is 1.08 V. This might be related to changes in opto-electronic properties obtained by deposition on different bottom layers. For (MA,FA,Cs)Pb(I/Br)$_3$ with and without K passivation, the dependencies of $\mu_F/e$ with intensity are very similar, while the measured $V_{oc}$ values differ significantly as shown in Figure 9b. This increase in $V_{oc}$ is attributed to the passivation by K of surface states formed by the deposition of the HTM layer. Hence, we suggest that the K-doping retards the interfacial recombination between the MHP and spiro-OMeTAD. Nonetheless, for both sets of samples shown in Figures 9a and 9b we can conclude that the measured $V_{oc}$ can reach values very close to $\mu_F/e$. Hence, by either optimizing the device structure and/or by effectively passivating the interfaces, considerable rise of the $V_{oc}$ can be realized. Further pushing the $V_{oc}$ to its theoretical limit of 1.33 V for MAPbI$_3$ should then be realized by increasing the $\mu_F/e$ in the perovskite layer, by reducing the non-radiative recombination.[70–72] Figure 9c shows a possible way to improve the $\mu_F/e$ by light soaking the perovskite film in humid air, yielding a $\mu_F/e$ value of 1.29 V. However, from the notable gap between $\mu_F/e$ and $V_{oc}$, we suggest that the device configuration was not optimal in that study, meaning that deposition of the transport layers enhances the non-radiative recombination. In fact, recent work demonstrates that a $V_{oc}$ of 1.26 V can be obtained by a proper selection of transport materials in combination with a perovskite layer prepared from PbAc$_2$ and PbCl$_2$ and using light-soaking as post-treatment.[69] This section shows that from the dynamic parameters obtained by fitting the TRMC data we can predict the concentrations of excess electrons and holes under continuous excitation. This allows us to calculate the Quasi-Femi level splitting, revealing the intrinsic limit of the $V_{oc}$ for a specific layer. This versatile methodology connects time-resolved fundamental research on charge carrier dynamics to practical device performance.

**Section 6: Outlook, limitations and possibilities**

Recent records of MHP solar cells are all based on mixed-cation/mixed-halide perovskites (MCMHPs), comprising formamidium (FA), MA and Cs as well as iodide and bromide. The advantages of MCMHPs are the tunable band-gaps and the more stable black phase making those materials suitable as top cell absorber material in a tandem solar cell.[78,79] However, the main drawback is that these materials are susceptible to light-induced phase segregation, yielding iodide rich domains.[80–82] Since these iodide-rich domains feature a smaller band-gap than the bulk of the MCMHPs layer these act as sinks to charge carriers, leading to effective collection of light-induced charge carriers.[83–85] The local high concentration of carriers leads to efficient radiative band-to-band recombination resulting in a substantially red-shifted PL. Apart from this light-induced phase segregation it is suggested that even before segregation these materials exhibit material inhomogeneities caused by well-intermixed halide distributions.[86] It is unclear to what extent the cations are involved with these properties. Figure 10 presents TRMC traces recorded on excitation of $(FA,MA,Cs)Pb(I_{0.6}Br_{0.4})_3$ before light-induced segregation. In particular from the log/lin representation mode it is clear that the TRMC tails exhibit the same slope independent of the laser intensity. This feature is not observed for MHPs with only a single halide such as $(FA,MA,Cs)PbI_3$ or for mixed-halide MHPs with only a single cation $MAPb(I_{0.6}Br_{0.4})_3$.[87] This implies that the occurrence of these TRMC tails is somehow linked to the presence of mixed cations and mixed halides.

The model as presented in Section 2 is not capable of describing the kinetics shown in Figure 10, implying that additional elements are required. For this specific example of $(FA,MA,Cs)Pb(I_{0.6}Br_{0.4})_3$ the introduction of shallow states was sufficient to observe a good match between experiment and model. Minority carriers get temporarily immobilized by these shallow states and are released back into the band by thermal motion. Temperature-dependent measurements showed that indeed at low temperatures a larger fraction of the charges remained in these shallow states preventing recombination by band-to-band recombination.

Although the traces in Figure 10 can be described satisfactorily, there are a number of remarks to be noted. The coupled differential equations discussed in Section 2 assume homogeneous charge carrier concentrations. For the materials presented and discussed in Section 3 this assumption is reasonable since by using different excitation sides or wavelengths no differences in decay kinetics are observed. This implies that within our experimental response time of a nanosecond, the light-induced charges are equally distributed throughout the entire MHP layer. However, in the MCMHPs featuring energetic inhomogeneities, charges rapidly localize in these low band-gap domains. This leads to underestimation of the band-to-band rate constant. To find accurate values, a better description of the domain sizes within the layer is required. Finally, we note that we assume that deep and shallow traps are homogenously distributed in the layer. In practice these states could be entirely located on the surface or on grain boundaries in the polycrystalline layer. To take this into account it would be required to include not only time, but also space as a variable, at least in one dimension. Such modelling has been done for MAPbBr$_3$ single crystals for which surface-mediated recombination is the most important decay process.[88] Such approach is very useful to come to an accurate description of the charge carrier dynamics and would give valuable insight which aspects in the perovskite are of importance to improve. Unfortunately, since the charge carrier mobilities are relatively high in these materials, the fast initial processes can not be resolved. Therefore, one of our main focus points for the near future is to improve the time resolution of the photoconductivity measurements, while maintaining the high sensitivity. Only in this way the conductivity measurements give enough information to come to a meaningful model. Furthermore, advanced models have to be developed which will teach us more about the band structure and intra-band gap levels of these more complex perovskite semiconductors.

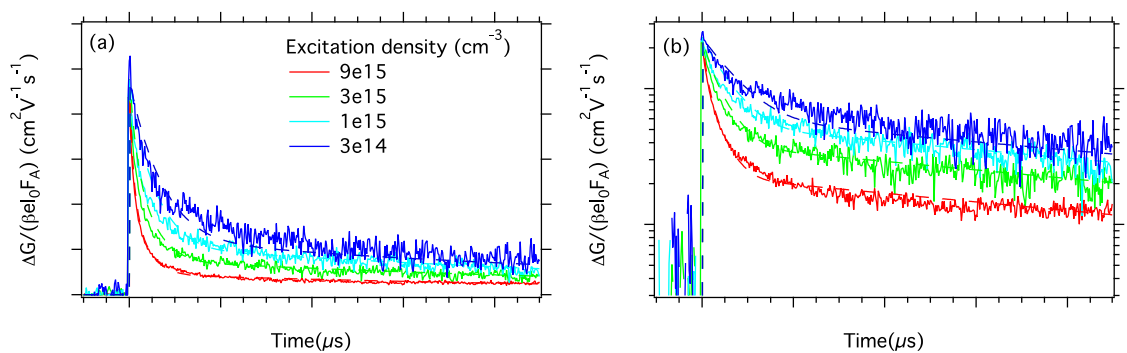

**Figure 10.** TRMC traces of (FA,MA,Cs)Pb(I$_{0.6}$Br$_{0.4}$)$_3$ observed on excitation at 500 nm leading to excitation densities as given in the annotations. The dashed lines are fits to the decay kinetics using the model presented in Section 2 complemented by additional shallow states. (b) gives the same information in a log/lin representation. Data from Guo et al. [87]